\documentclass[aps,prl,twocolumn,showpacs,groupedaddress,amsmath,amssymb]{revtex4}

\usepackage{graphicx}
\usepackage{color}
\usepackage{comment}
\usepackage{bm}
\usepackage{latexsym}

\begin{document}

\title{Floquet Scattering Theory based on Effective Hamiltonians of Driven Systems}
\author{Huanan Li$^1$, Boris Shapiro$^2$, Tsampikos Kottos$^1$} 
\affiliation{$^1$Wave Transport in Complex Systems Lab, Department of Physics, Wesleyan University, Middletown, CT-06459, USA}
\affiliation{$^2$Technion - Israel Institute of Technology, Technion City, Haifa 32000, Israel}
\date{\today}

\begin{abstract}
The design of {\it time-independent effective Hamiltonians} that describe periodically modulated systems, provides a promising approach 
to realize new forms of matter. This, so-called, Floquet engineering approach is currently limited to the description of wavepacket dynamics. 
Here, we utilize the notion of effective Hamiltonians and develop a Floquet engineering scattering formalism that relies on a systematic 
high-frequency expansion of the scattering matrix. The method unveils the critical role of micromotion. An application to the case of non-reciprocal transport is presented. 
\end{abstract}

\pacs{05.45.-a, 42.25.Bs, 11.30.Er}
\maketitle
{\it Introduction -- }The design of periodically driven schemes and their implementation to various physical systems has attracted the attention 
of the research community during the last decade \cite{BDP15,E17,ANG17}. The interest in this activity, coined {\it Floquet engineering}, is twofold: 
from the fundamental side one hopes that these new developments will allow for the realization of novel phenomena and forms of matter 
which are currently out of reach when using conventional (material development-based) approaches. Examples include systems with dynamical 
localization \cite{DK86,MRBWR94,MFDNR98,SGHSDPNTLK10,LKS18a,Segev18}, topologically nontrivial phases \cite{LRG11,HTCOSSWWSLE12}, 
quantum phase transitions \cite{ZLCMA09,EH07}, artificial gauge potentials \cite{GD14,ANG17}, edge states \cite{GP13,RLBL13} etc. From the 
technological side, the hope is that these advances can be utilized towards the development of novel devices that will provide unprecedented
control of information carried over by a variety of systems, ranging from optical, microwave, and acoustic to matter waves and quantum electronic 
framework. This enthusiasm is further backed up by the fact that the properties of a Floquet system are reconfigurable and can be changed on
demand by altering the external driving field -- a feature that is absent from any ``time-independent" system whose physical properties are 
fixed during the fabrication process. 

The cornerstone concept in the Floquet engineering approach is associated with the notion of effective (Floquet) Hamiltonian $\hat{H}_{\rm F}$. 
The latter is {\it time-independent} and allows us to describe the evolution of the original periodically driven system $\hat{H}_0(t+T)=\hat{H}_0(t)$ 
in a stroboscopic manner. In other words, the evolution of the driven system is given by $\hat{U}\left(t_{2},t_{1}\right)=\hat{U}_{F}\left(t_{2}\right)
e^{-\imath\left(t_{2}-t_{1}\right)\hat{H}_{F}}\hat{U}_{F}^{\dagger}\left(t_{1}\right)$ where $\hat{U}_F(t)=\hat{U}_F(t+T)$ encodes the dynamics taking 
place within each period $T$ of the drive, the so-called micromotion. The prerequisite for Floquet engineering is that $\hat{H}_F$ can be designed 
by suitably tailoring the driving protocol \cite{GD14,BDP15,RGF03,GDAC15,EA15}. Of course, the implicit assumption is that one can theoretically 
calculate the effective Hamiltonian (as well as the micromotion operator) associated with the specific driving scheme. In general, this is a formidable 
task. There are, however, experimentally relevant circumstances where $\hat{H}_F$ (and 
also $\hat{U}_F(t)$) can be computed in the form of an inverse-frequency expansion \cite{GD14,RGF03,EA15} and takes a simple 
form that allows for a clear interpretation. In this fast driving regime, in which the driving frequency is larger than any natural energy scale in the 
problem, the slow degrees of freedom are not coupled resonantly with the drive. Instead, the system typically feels an effective static potential 
that depends on the amplitude and frequency of the drive and which can be described by $\hat{H}_F$. Experimental implementation of this scheme 
includes, dynamical trapping of Paul traps \cite{LBMW03} and dynamical localization \cite{MRBWR94,MFDNR98,SGHSDPNTLK10}, to artificial 
gauge fields for neutral atoms \cite{JMDLUGE14,KBCK15,ALSABNCBG15}, nontrivial topological band structures \cite{GBZ16} and quantum phase 
(superfluid-to-Mott) transitions \cite{EWH05,ZLCMA09,EH07}. It is important to point out that all the existing studies on Floquet engineering are 
mainly concerned with the so-called wavepacket dynamics scenario (i.e. the spreading of an excitation). It turns out that all these studies 
highlight the importance of $\hat{H}_F$, while the micromotion $\hat{U}_F(t)$ appears to play a secondary (if at all) role in the description of the 
dynamics.

\begin{figure}
\includegraphics[width=1\linewidth, angle=0]{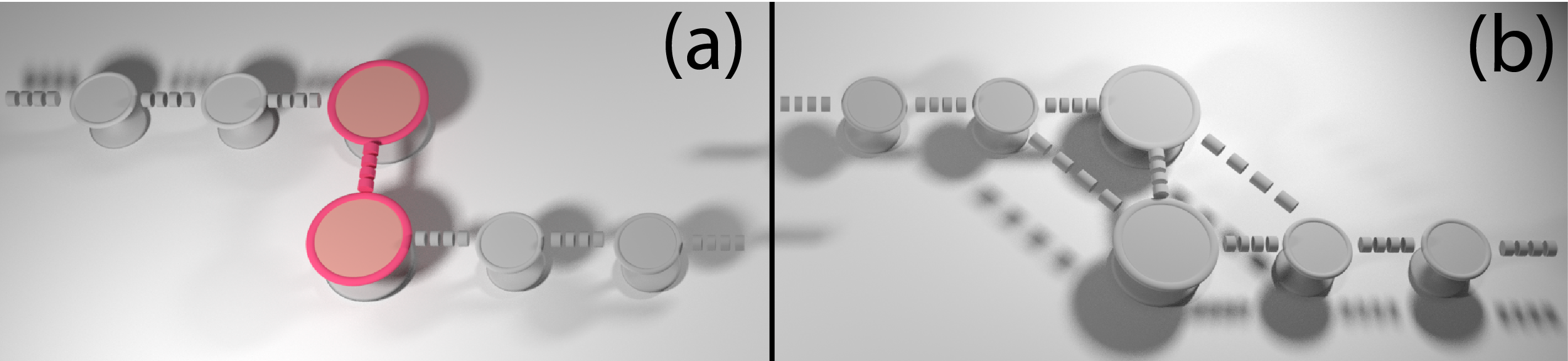}
\caption{(a) A schematic of a Floquet two-level system (two coupled resonators-- red color) coupled to two semi-infinite leads. The resonance frequency,
and the mutual coupling of the (red) resonators, can be periodically modulated in time; (b) The same system
can be described effectively as two time-independent resonators for which the on-site potential, their mutual coupling and their coupling to the leads 
are now altered by the periodic drive. This system, in the scattering framework, is now described by an effective time-independent Hamiltonian 
$\hat{H}_F$. Note that up to order $O(1/\omega^2)$ the alteration of the coupling with the leads is solely dictated by the micromotion.}
\label{fig1}
\end{figure}

Besides all this activity, however, not much has been done for the implementation of the effective (Floquet) Hamiltonian notion within the scattering 
framework. This is quite surprising since the latter is the appropriate formalism for the analysis of the transport properties of a system. 
In this paper, we undertake this task and develop a {\it Floquet engineering scattering theory based on effective Hamiltonians of periodically driven 
systems}. Specifically we develop a systematic high-frequency expansion of the Floquet scattering matrix, which unveils the important role of 
micromotion in modifying the coupling of the system to the leads. We demonstrate the efficiency of our scheme, by utilizing it for the design of periodic 
driving protocols that aim to enhanced (optimized) non-reciprocal transport (NRT).

{\it Theoretical Formalism--}We consider a Floquet system described by a finite-dimensional $N_s\times N_s$ time-periodic Hamiltonian $\hat{H}
\left(t\right)=\hat{H}\left(t+T\right)$ ($T\equiv2\pi/\omega$ is the period of the drive and $\omega$ is its frequency),
\begin{align}
\hat{H}\left(t\right) & =\hat{H}_{0}\left(t\right)-\imath\hat{\Gamma};\quad\hat{\Gamma}=\gamma I_{N_{s}},
\label{eq1}
\end{align}
where $\hat{H}_{0}\left(t\right)$ is a $N_{s}\times N_{s}$ hermitian matrix, $-\imath\hat{\Gamma}$ describes the uniform losses in the system
and $I_{N}$ (with an arbitrary subscript $N$) denotes the $N\times N$ identity matrix. For example, in the context of the coupled mode
theory, the Hamiltonian $\hat{H}\left(t\right)$ could describe a network of coupled single-mode resonators (sites) with uniform losses $\gamma$, 
whose resonant frequencies and couplings are modulated periodically in time, see Fig.~\ref{fig1}a. We turn the system of Eq.~(\ref{eq1}) to a scattering set-up 
by attaching to it two static semi-infinite leads $\alpha=L,R$, each of which is supporting plane waves with a dispersion relation $E\left(k\right)$.

For periodic driving, the Floquet theorem assures that when an incident wave with frequency $E_0=E(k_0)$ is approaching the time-periodic modulated 
target, it will scatter to infinite number of outgoing channels (including evanescent channels) with frequencies $E_{n}\left(k_{n}\right)=E_0+n\omega$ 
where $n$ is an integer. We are interested in the scattering matrix $\mathcal{S}$, which connects only the propagating incoming and outgoing 
channels. Following the derivation in Ref.~\cite{Li2018}, we can calculate the flux-normalized $2N_{p}\times2N_{p}$ scattering matrix $\mathcal{S}$
where $N_{p}$ is the number of propagating channels in each lead. We have
\begin{align}
\mathcal{S} =-I_{2N_{p}}+\imath W\frac{1}{E_0+\imath\gamma-H_{Q}+\Lambda_H+\frac{\imath}{2}W^{T}W}W^{T}\label{eq: scattering_matrix}
\end{align}
where $H_{Q}=H_{Q}^{0}-\left[n\omega I_{N_{s}}\right]$ is the quasi-energy operator in the extended Floquet-Hilbert space, with the matrix 
component $\left(H_{Q}^{0}\right)_{ns,n's'}=\hat{H}_{ss'}^{\left(n-n'\right)}\equiv\frac{1}{T}\int_{0}^{T}dt\left(\hat{H}_{0}\left(t\right)\right)_{ss'}
e^{\imath\left(n-n'\right)\omega t}$. The subindex 
$s=1,\cdots, N_s$ indicates the $s$-site of the time-periodic system. The notation $\left[X_{n}\right]$ denotes an infinite block diagonal matrix with each diagonal block being 
a submatrix $X_{n}$ (the integer $n$ runs from $-\infty$ to $+\infty$). The hermitian operator $\Lambda_H = W_{c}^{T}\left[\varLambda_{n}I_{2}\right]W_{c}$ describes the channel-coupling induced renormalization for the ``Hamiltonian'' $H_{Q}$ in the extended Floquet-Hilbert space.  Here the  
matrix elements $\varLambda_n$ distinguish between propagating and evanescent channels. Furthermore the operator $\left(W_{c}\right)_{n\alpha,n's}=c_{\alpha}\delta_{nn'}
\delta_{\alpha\leftrightarrow s}$ describes the coupling between the channels and the sites of the driven system in the Floquet-Hilbert space,  where $c_{\alpha}$ is the bare coupling between the lead $\alpha$ and the sites of the system, and we define $\delta_{\alpha\leftrightarrow s}=1$ when 
the lead $\alpha$ is directly coupled with the site $s$ and $\delta_{\alpha\leftrightarrow s}=0$ otherwise. Finally $\left(W\right)_{n_{P}\alpha,n's}
=\sqrt{v_{g,n_{P}}}c_{\alpha}\delta_{n_{P}n'}\delta_{\alpha\leftrightarrow s}$ describes the coupling of  the propagating channels with the driven system in the extended space. The integer $n_{P}$ labels only the propagating channels and $v_{g,n_{P}}=\left.\partial E/\partial k\right|_{k_{n_{P}}}$
represents the group velocity in each of these channels.

We proceed with the high-frequency analysis of the scattering matrix Eq.~(\ref{eq: scattering_matrix}). Our goal is to incorporate the effective 
Hamiltonian $\hat{H}_{F}$ in the Floquet scattering framework. Indeed, $\hat{H}_{F}$ 
is closely connected with the quasi-energy operator $H_{Q}$ as \cite{EA15}
\begin{align}
U_{F}^{\dagger}H_{Q}U_{F} & =\left[\hat{H}_{F}\right]-\left[n\omega I_{N_{s}}\right],U_{F}^{\dagger}U_{F}=\left[I_{N_{s}}\right],
\label{eq: diagonalization}
\end{align}
where the sub-blocks $\hat{u}_{n-n'}\equiv\left(U_{F}\right)_{n,n'}$ of the unitary matrix $U_{F}$ are related to the one-point micromotion operator
$\hat{U}_{F}\left(t\right)$ as $\hat{u}_{n}=\frac{1}{T}\int_{0}^{T}dt\hat{U}_{F}\left(t\right)e^{\imath n\omega t}$. Therefore, $U_{F}$ 
encodes the micromotion generated by $\hat{H}_{0}\left(t\right)$ or equivalently $\hat{H}\left(t\right)$ in the uniform-loss case.
Note that in general, both the effective Hamiltonian $\hat{H}_{F}$ and the matrix $U_{F}$ are not uniquely defined. However, under the 
canonical van Vleck form of degenerate perturbation theory, they can be uniquely determined order by order with respect to $1/\omega$ 
in terms of the Fourier components $\hat{H}^{\left(n\right)}$ of the $\hat{H}_{0}\left(t\right)$, see Ref.~\cite{EA15} and \cite{note1}. We note that 
the off-diagonal blocks of $U_F$, i.e.  $\hat{u}_{n\neq 0}$, have been found to be $\sim O(1/\omega)$ \cite{EA15}. 

In the large driving-frequency limit, the number of the propagating channels in each lead reduce to one, i.e., $N_{p}=1$. Using Eqs.~(\ref{eq: scattering_matrix}) 
and (\ref{eq: diagonalization}), we can write the scattering matrix $\mathcal{S}$ in terms of the effective Hamiltonian $\hat{H}_{F}$ as 
\begin{align}
S & =-I_{2}+\imath W^{F}G_{s}^{F}W^{F\dagger}\label{eq: S_2by2}
\end{align}
where $W^{F}=WU_{F}$ and
\begin{align}
G_{s}^{F} & =\frac{1}{\left[\left(E_{n}+\imath\gamma\right)I_{2}\right]-\left[\hat{H}_{F}\right]+U_F^{\dagger}\left(\Lambda_H+\frac{\imath}{2}W^TW\right)U_F}.
\end{align}
The difficulty of the analysis of the scattering matrix $\mathcal{S}$ mainly comes from the presence of the full matrix $U_F^{\dagger}\left(\Lambda_H+\frac{\imath}{2}W^TW\right)U_F$ in $G_{s}^{F}$. The above matrix can be formally separated into a block-diagonal part $\left[\hat{d}_{n}\right]$ and a block-off-diagonal matrix
$V$, i.e., 
\begin{align}
U_F^{\dagger}\left(\Lambda_H+\frac{\imath}{2}W^TW\right)U_F & =\left[\hat{d}_{n}\right]+V
\label{decomp}
\end{align}
where $V_{n_{1},n_{2}}=\hat{d}_{n_{1,}n_{2}}$ when $n_{1}\neq n_{2}$ and $\hat{d}_{n}=\hat{d}_{n,n}$. The above decomposition Eq.~(\ref{decomp}) to
a block-diagonal and a block off-diagonal matrices is inspired by a locator expansion method used in disordered systems \cite{Eco76}. Though, in the current
situation, the locator is a matrix. Specifically, taking into account that $V\sim O\left(1/\omega\right)$, the matrix $G_{s}^{F}$ can be expanded systematically 
as 
\begin{align}
G_{s}^{F} & =-\left(g^{F}+g^{F}Vg^{F}+\cdots\right),\label{eq: expansion of GsF}
\end{align}
where  $g^{F}=\left[\hat{g}_{n}^{F}\right]$ and the ``locator'' is identified as  $\hat{g}_{n}^{F}=-\left(E_{n}+\imath\gamma-\hat{H}_{F}+\hat{d}_{n}\right)^{-1}$.
Notice that for $n\neq 0$ the locator is of order $g_{n\neq0}^{F}\sim O\left(1/\omega\right)$. 

Our final step involves the expansion of $W^{F}=WU_{F}$ (specifically of $U_F$) in powers of $1/\omega$ and its substitution, together with 
Eq.~(\ref{eq: expansion of GsF}), into the expression of the $S$-matrix Eq. (\ref{eq: S_2by2}). In this way, a systematic high-frequency expansion 
for the scattering matrix $\mathcal{S}$ can be readily derived.

Next, let us consider a scenario for which $\varLambda_{n\neq 0}\sim O(1/\omega)$. This is a typical situation due to the diminishing coupling between 
the driving system and the evanescent channels $E_{n\neq 0}$. Up to order $O\left(1/\omega^{2}\right)$, we can write the scattering matrix $\mathcal{S}$ 
in Eq.~(\ref{eq: S_2by2}) as 
\begin{align}
S & \approx-I_{2}+\imath \hat{c}\hat{u}_{0}\frac{v_{g,0}}{E_0+\imath\gamma-\hat{H}_{F}+\hat{u}_{0}^{\dagger}\hat{c}^{\dagger}
\left(\varLambda_0+\imath{v_{g,0}\over2}\right)\hat{c}\hat{u}_{0}}\hat{u}_{0}^{\dagger}\hat{c}^{\dagger},\label{eq: S_high_w}
\end{align}
where $\left(\hat{c}\right)_{\alpha s}=c_{\alpha}\delta_{\alpha\leftrightarrow s}$, $\hat{H}_{F}=\hat{H}_{F}^{\dagger}\approx\hat{F}_{0}+
\frac{1}{\omega}\hat{F}_{1}+\frac{1}{\omega^{2}}\hat{F}_{2}$ and $\hat{u}_{0}\approx I_{2}+\frac{1}{\omega^{2}}\hat{b}_{2}$. The operators 
$\hat{F}_n$ appearing above are
\begin{align}
\hat{F}_{0} & =\hat{H}^{\left(0\right)},\quad\hat{F}_{1}=-\sum_{m=1}^{\infty}\frac{1}{m}\left[\hat{H}^{\left(m\right)},\hat{H}^{\left(-m\right)}\right],\\
\hat{F}_{2} & =\sum_{m\neq0}\frac{\left[\hat{H}^{\left(-m\right)},\left[\hat{H}^{\left(0\right)},\hat{H}^{\left(m\right)}\right]\right]}{2m^{2}}\nonumber \\
 & +\sum_{m\neq0}\sum_{m'\neq0,m}\frac{\left[\hat{H}^{\left(-m'\right)},\left[\hat{H}^{\left(m'-m\right)},\hat{H}^{\left(m\right)}\right]\right]}{3mm'}\nonumber 
\end{align}
and $\hat{b}_{2}=-\frac{1}{2}\sum_{m\neq0}\frac{1}{m^{2}}\hat{H}^{\left(-m\right)}\hat{H}^{\left(m\right)}$ \cite{note2}. From Eq.~(\ref{eq: S_high_w}) 
we see clearly that the micromotion via $\hat{u}_{0}$ is modifying the coupling of the driven system to the leads, see for example Fig.~\ref{fig1}b. 
Notice that the scattering matrix $S$ in Eq.~(\ref{eq: S_high_w}) is independent of the driving phase, i.e., $S$ is invariant when
$\hat{H}\left(t\right)\rightarrow\hat{H}\left(t-t_{0}\right)$. This property is guaranteed by the structure of both $\hat{H}_{F}$ and
$\hat{u}_{0}$, which involve only the matrix products $\hat{H}^{\left(m_{1}\right)}\hat{H}^{\left(m_{2}\right)}\cdots\hat{H}^{\left(m_{p}\right)}$
with $m_{1}+m_{2}+\cdots m_{p}=0$. Generally, the leading correction of the micromotion to the Floquet scattering problem 
is ${\cal O}(1/\omega^2)$. This is the same order where novel topological properties of ${\hat H}_F$ of the closed system typically manifest themselves. 
The correct implementation of ${\hat H}_F$ in the scattering framework (where the micromotion is taken into account) is 
crucial in order to avoid any misinterpretations.

For example, in the case of a tight-binding lead with dispersion relation $E\left(k\right)=-2\cos(k)$ (in units of coupling), Eq. (\ref{eq: S_high_w}) applies
with $v_{g,0}=2\sin(k_0)$. In this case $\varLambda_{n}=\cos (k_{n}),k_{n}\in\left(0,\pi\right)$ for the propagating channels $E_{n}\in
\left(-2,2\right)$ and $\varLambda_{n}=e^{\imath k_{n}}\sim O\left(1/\omega\right), {\cal I}m\left(k_{n}\right)>0$ for the evanescent channels $E_{n}\notin
\left(-2,2\right)$. Therefore we have that $\hat{d}_{n_{1,}n_{2}}=\sum_{n}\hat{u}_{n-n_{1}}^{\dagger}\hat{c}^{\dagger}e^{\imath k_{n}}\hat{c}
\hat{u}_{n-n_{2}}$ (see Eq. (\ref{decomp})).

{\it Application: Design of driving schemes for enhanced NRT --} Let us now implement the above formalism for the Floquet design of driving
schemes that can lead to Non-Recirpocal Transport (NRT). We will confine our analysis to the high-frequency limit where 
Eq.~(\ref{eq: S_high_w}) is applicable. The corresponding physical requirement is that the driving amplitudes are small compared with the 
driving frequency $\omega$. Furthermore, we will consider that the scattering target is having uniform losses of strength $\gamma$. In this 
case, the scattering matrix $S$ is not unitary i.e. $I_{2}-S^{\dagger}S=\gamma Z$ where $Z=Z^{\dagger}$ is the absorption matrix. We remark, 
parenthetically, that one can utilize this framework in order to design driving schemes that lead to a broadband absorption \cite{LSK18}.

We proceed with the evaluation of the left/right transmittance difference $\Delta\equiv T_{L\rightarrow R}-T_{R\rightarrow L}$, where 
$T_{L\rightarrow R}=\left|t_{L}\right|^{2}$ represents the transmittance from left to right lead for the propagating channel $E_{0}$ (and 
similarly for $T_{R\rightarrow L}=\left|t_{R}\right|^{2}$) . We get
\begin{align}
\Delta & =\gamma\mathrm{Tr}\left(Z\begin{pmatrix}t_{R} & 0\\
0 & t_{L}
\end{pmatrix}S^{*}\begin{pmatrix}0 & 1\\
-1 & 0
\end{pmatrix}\right), S=\begin{pmatrix}r_{L} & t_{R}\\
t_{L} & r_{R}
\end{pmatrix}.
\label{eq: transmittance_diff}
\end{align}
Equation (\ref{eq: transmittance_diff}) applies to any periodically driven system coupled to two propagating channels. 
We point out that Eq. (10) implies that the transmittance asymmetry and thus NRT will be absent without material loss i.e., 
$\gamma=0$. Indeed, this fact is the direct result of a $2\times2$ unitary scattering matrix. Further substitution of the 
expression Eq. (\ref{eq: S_high_w}) for the scattering matrix into Eq. (\ref{eq: transmittance_diff}) will allow us to optimize 
the driving scheme that will lead to maximum NRT. 

Let us consider, for demonstration purposes, the example case of a simple two-mode Floquet system corresponding to $N_{s}=2$. We will
further assume that the Floquet system is coupled to two tight-binding leads. In this situation 
the bare coupling matrix takes the form $\hat{c}=\begin{pmatrix}c_{L} & 0\\0 & c_{R}\end{pmatrix}$. For simplicity, we assume that $c_{L}=c_{R}=-1$ 
and the static part of $\hat{H}_{0}\left(t\right)$ is $\hat{F}_{0}=\begin{pmatrix}0 & h_{0}\\h_{0} & 0\end{pmatrix},h_{0}\in\mathcal{R}$. Under this 
simplification, the parametrization $\hat{b}_{2}=\hat{b}_{2}^{\dagger}=\begin{pmatrix}\beta_{1} & \beta_{r}+\imath\beta_{i}\\
\beta_{r}-\imath\beta_{i} & \beta_{2}
\end{pmatrix}$ together with Eqs.~(\ref{eq: S_high_w}) and (\ref{eq: transmittance_diff})
enable us to obtain the transmittance difference $\Delta$ up to the
order $O\left(1/\omega^{2}\right)$ explicitly 
\begin{align}
\Delta & \approx\frac{(32h_{0}\beta_{i}\gamma \sin^{2}k_{0})/\omega^{2}}{4\left(\gamma+\sin k_{0}\right)^{2}\cos^{2}k_{0}+
\left(h_{0}^{2}-\cos^{2}k_{0}+\left(\gamma+\sin k_{0}\right)^{2}\right)^{2}}.
\label{eq: Delta}
\end{align}
From Eq.~(\ref{eq: Delta}) we conclude that in this case, the leading order of NRT in the high-frequency limit is associated with the 
presence of micromotion and it is $O\left(1/\omega^{2}\right)$. In fact, this estimation can be derived directly from Eq.~(\ref{eq: S_high_w}), 
for a generic (i.e. arbitrary $\hat{F}_0=\hat{F}_0^{\dagger}$ and $c_L, c_R$) two-mode Floquet system. This result stresses the
importance of the micromotion in the scattering framework -- its overlook can lead to qualitatively wrong conclusions, as the example case 
of non-reciprocal transport indicates.

We proceed with the Floquet engineering design of NRT, under the constraint that the driving scheme involves only a first-harmonic i.e. 
\begin{align}
\hat{H}_{0}\left(t\right) & =\begin{pmatrix}\varepsilon_{L}\left(t\right) & h\left(t\right)\\
h^{*}\left(t\right) & \varepsilon_{R}\left(t\right)
\end{pmatrix}\label{eq: example_H0}
\end{align}
where $\varepsilon_{L/R}\left(t\right)=2f_{L/R}\cos\left(\omega t+\phi_{L/R}\right)$ and $h\left(t\right)=h_{0}+h_{1}\exp\left(\imath\omega t+\imath\phi_{1}\right)+h_{2}\exp\left(-\imath\omega t-\imath\phi_{2}\right)$. Without loss of generality, we set $\phi_{L}=0$ (we can always 
choose the driving phase properly to eliminate $\phi_{L}$ without affecting NRT). In this respect, our goal is to maximize the transmittance 
difference $\Delta$ by choosing the proper \textit{relative} driving phases, i.e., $\phi_{R}$, $\phi_{1}$ and $\phi_{2}$ when $\phi_{L}=0$. 
Therefore the problem boils down to the study of $\beta_{i}$ appearing in Eq.~(\ref{eq: Delta}). We get 
\begin{align}
\beta_{i} & =\frac{1}{2}\Big(-f_{L}h_{1}\sin\phi_{1}+f_{L}h_{2}\sin\phi_{2}\nonumber \\
 & -f_{R}h_{1}\sin\left(\phi_{1}-\phi_{R}\right)+f_{R}h_{2}\sin\left(\phi_{2}-\phi_{R}\right)\Big).\label{eq: beta_i}
\end{align}
From Eqs.~(\ref{eq: Delta},\ref{eq: beta_i}) we find that in case of real-valued coupling $h\left(t\right)\in R$ that involves only a first-harmonic 
driving associated with the example of Eq.~(\ref{eq: example_H0}), the transmittance asymmetry cannot be stronger than $O\left(1/
\omega^{3}\right)$. On the other hand, when the coupling is assumed to take complex values, we have that $\max\{\beta_{i}\}=\frac{1}{2}
\left(f_{L}h_{1}+f_{L}h_{2}+f_{R}h_{1}+f_{R}h_{2}\right)$ occurs when $\phi_{R}=0$, $\phi_{2}=-\phi_{1}=\pi/2$, thus resulting in an optimized NRT. 

\begin{figure}
\includegraphics[width=1\linewidth, angle=0]{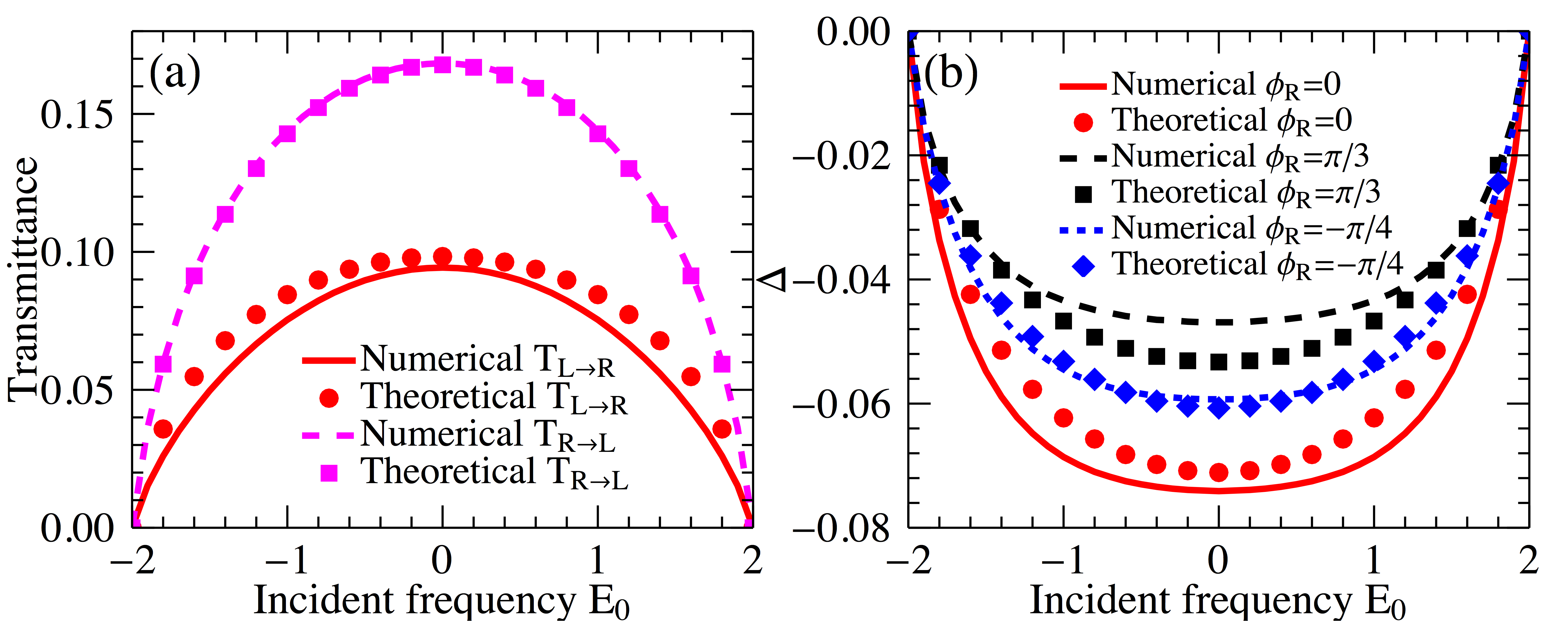}
\caption{The analysis of transport properties of the specific two-mode Floquet
system defined in Eq.~(\ref{eq: example_H0}). (a) Numerical and
theoretical calculation of $T_{L\rightarrow R}$ ($T_{R\rightarrow L}$)
for the transmittances from the left (right) to right (left) lead
versus the incident frequency $E_0$ in the case of the optimized phase
when $\phi_{L}=\phi_{R}=0$ and $\phi_{2}=-\phi_{1}=\pi/2$. Theoretical
calculation refers to the application of Eq.~(\ref{eq: S_high_w})
while the numerical rigorous results come from Eq.~(\ref{eq: scattering_matrix})
and the convergence was tested by increasing the Floquet lattice.
(b) The transmittance difference $\Delta$ versus the incident frequency
$E_0$ for various different relative driving phases $\phi_{R}=0,$
$\pi/3$ and $-\pi/4$ when $\phi_{L}=0$ and $\phi_{2}=-\phi_{1}=\pi/2$.
Theoretical results come from the explicit formula Eq.~(\ref{eq: Delta})
with $\beta_{i}$ given in Eq.~(\ref{eq: beta_i}). Clearly the optimized
phase $\phi_{R}=0$ presents a relatively large magnitude of the transmittance
difference $\left|\Delta\right|$. The other common parameters for
the calculation of both (a) and (b) are: the bare coupling between
the leads and the Floquet system $c_{L}=c_{R}=-1$, the static coupling
between the two modes $h_{0}=-1$, the driving amplitudes $f_{L}=f_{R}=h_{1}=h_{2}=1$
and the uniform loss strength $\gamma=1$ and the driving frequency
$\omega=6$.\label{fig: transmittance}}
\end{figure}

In Fig.~\ref{fig: transmittance}a, we present the numerical (lines) results for the left ($T_{L\rightarrow R}$) and right ($T_{R\rightarrow L}$) 
transmittance versus the 
frequency of an incident wave for the Floquet system of Eq.~(\ref{eq: example_H0}). In our simulations we have used high values of the 
driving frequency $\omega=6$ and the optimized values of the relative driving phases which we have find to be $\phi_{L}=\phi_{R}=0$ 
and $\phi_{2}=-\phi_{1}=\pi/2$. At the same figure we also present the theoretical (symbols) results for $T_{L\rightarrow R}$ and $T_{R
\rightarrow L}$ which are based on Eq. (\ref{eq: S_high_w}). Finally in Fig. \ref{fig: transmittance}b we show the theoretical 
(symbols) and numerical (lines) transmittance difference $\Delta$ for various values of the relative driving phases $\phi_{R}$. 
The maximum magnitude of the transmittance difference is observed for $\phi_{R}=0$ as predicted by our theoretical analysis.

{\it Conclusions --} We have developed a scattering Floquet engineering approach which is applicable in the high driving-frequency limit.
The method utilizes the notion of the effective (Floquet) Hamiltonian which have been recently developed in the framework of wavepacket 
dynamics of Floquet systems. The approach highlights the importance of micromotion in the scattering framework and allows us to design
driving schemes with predefined transport characteristics. We have demonstrated the validity of our scheme by applying it to the case of 
non-reciprocal transport associated with a two-site periodically modulated target with uniform losses and a first-harmonic time-periodic modulation. 
We furthermore find that the transmittance asymmetry is proportional to $~O(1/\omega^2)$ for any two-site Floquet-system with uniform
loss. It will be interesting to utilize the scattering Floquet engineering approach in order 
to identify periodically modulated scattering geometries for which the non-reciprocity will be of order $O\left(1/\omega\right)$. A promising 
direction along this lines is the investigation of modulated targets with spatio-temporal symmetries (like parity-time symmetry) \cite{CLEK17,LSK18}. 

{\it Acknowledgments-- } 
This research was partially supported by DARPA NLM program via grant No. HR00111820042, by an AFOSR Grant No. FA 9550-10-1-0433, and by NSF Grant No. EFMA- 1641109.
B.S. acknowledges the hospitality of the Physics Department of Wesleyan University, where this work was performed. The views and opinions expressed in this paper are those of the authors and do not reflect the official policy or position of the U.S. Government.


\begin{thebibliography}{1}
\bibitem{note1}We remark that our convention is different from Ref.~\cite{EA15}.
In order to get consistent formula in our case, one needs to perform
$\omega\rightarrow-\omega$ for the corresponding ones in Ref.~\cite{EA15}.
\bibitem{note2}The notation $\left[\hat{A},\hat{B}\right]=\hat{A}\hat{B}-\hat{B}\hat{A}$
represents the standard commutator, which should not be confused with
the notation $\left[X_{n}\right]$ we introduced before.
\end{thebibliography}
\end{document}